\documentclass[12pt,article]{article}
\usepackage{natbib,graphicx}

\begin{document}

{\it
\noindent
Melnikov A.V., Shevchenko I.I.\\
``The rotation states predominant among the planetary satellites''\\
Icarus. 2010. V.209. N2. P.786--794.
}

\vskip2cm

\begin{center}
{\bf \Large The rotation states predominant among the planetary satellites}\\

\bigskip

{\it A. V. Melnikov\, and I. I. Shevchenko \\
Pulkovo Observatory of the Russian Academy of Sciences,\\
Pulkovskoje ave. 65/1, St.Petersburg 196140, Russia}

\end{center} 

\begin{abstract}

On the basis of tidal despinning timescale arguments, Peale showed
in 1977 that the majority of irregular satellites (with unknown
rotation states) are expected to reside close to their initial
(fast) rotation states. Here we investigate the problem of
the current typical rotation states among all known
satellites from a viewpoint of dynamical stability. We explore
location of the known planetary satellites on the
($\omega_0$, $e$) stability diagram, where $\omega_0$ is an
inertial parameter of a satellite and $e$ is its orbital
eccentricity. We show that most of the satellites with unknown
rotation states cannot rotate synchronously, because no stable
synchronous 1:1 spin-orbit state exists for them. They rotate
either much faster than synchronously (those tidally unevolved)
or, what is much less probable, chaotically (tidally evolved
objects or captured slow rotators).

\end{abstract}

\section{Introduction}
\label{sec:intro}

What is a typical rotation state of a planetary satellite? The
majority of planetary satellites with known rotation states rotate
synchronously (like the Moon, facing one side towards a planet),
i.e., they move in synchronous spin-orbit resonance 1:1. The data
of the NASA reference guide~\citep{NASA09} combined with
additional data (on the rotation of Caliban~(U16), Sycorax~(U17)
and Prospero~(U18) \citep{Maris01, Maris07} and the rotation of
Nereid~(N2)~\citep{Grav03}) implies that, of the 33 satellites
with known rotation periods, 25 rotate synchronously.

For the tidally evolved satellites, this observational fact is
theoretically expected. Synchronous 1:1 resonance with the
orbital motion is the most likely final mode of the long-term
tidal evolution of the rotational motion of a planetary satellite
\citep{GP66, P69, P77}.

Another qualitative kind of rotation known from observations is
fast regular rotation. There are seven satellites known to
rotate so~\citep{NASA09, Maris01, Maris07, Grav03, Bauer04}:
Himalia~(J6), Elara~(J7), Phoebe~(S9), Caliban~(U16),
Sycorax~(U17), Prospero~(U18), and Nereid~(N2); all of them are
irregular satellites (with a possible exception of Nereid, see
\cite{SJK06} and references therein). The rotation periods of them
are equal to $0.4$, $0.5$, $0.4$, $0.11$, $0.15$, $0.19$ and
$0.48$ days, respectively; i.e., they are less than their orbital
periods approximately 630, 520, 1400, 5200, 8600, 10300 and 750
times, respectively. These satellites, apparently, are tidally
unevolved.

A third observationally discovered qualitative kind of rotation is
chaotic tumbling. \citet{WPM84} and \citet{W87} demonstrated
theoretically that a planetary satellite of irregular shape in an
elliptic orbit could rotate in a chaotic, unpredictable way. They
found that a unique (at that time) probable candidate for the
chaotic rotation, due to a pronounced shape asymmetry and
significant orbital eccentricity, was Hyperion~(S7). Besides, it
has a small enough theoretical timescale of tidal deceleration of
rotation from a primordial rotation state. Later on, a direct
modelling of its observed light curves~\citep{K89b, BNT95, D02}
confirmed the chaotic character of Hyperion's rotation. Recent
direct imaging from the {\it CASSINI} spacecraft supports these
conclusions~\citep{T07}.

It was found in a theoretical research~\citep{KS05} that two other
Saturnian satellites, Pro\-me\-theus~(S16) and Pandora~(S17),
could also rotate chaotically (see also \citet{MS08}). Contrary to
the case of Hyperion, possible chaos in rotation of these two
satellites is due to occasional fine-tuning of the dynamical and
physical parameters rather than to a large extent of a chaotic
zone in the rotational phase space.

We see that the satellites spinning fast or tumbling chaotically
are a definite minority among the satellites with known rotation
states. However, the observed dominance of synchronous behaviour
might be a selection effect, exaggerating the abundance of the
mode typical for big satellites. This is most probable.
\citet{P77} showed on the basis of tidal despinning timescale
arguments that the majority of the irregular satellites are
expected to reside close to their initial (fast) rotation states.

A lot of new satellites has been discovered during last years. Now
the total number of satellites exceeds 160~\citep{NASA09}. The
rotation states for the majority of them are not known. All small
enough satellites have irregular shapes (and many of them large
orbital eccentricities, see \cite{Sh06}), and this may result, as
in the case of Hyperion~\citep{WPM84}, in the non-existence of
attitude stable synchronous state; or, such a state may be even
absent in the present epoch in the phase space of the planar
rotational motion.

In this paper, we investigate the problem of the current
typical rotation states among satellites from a viewpoint of
dynamical stability, considering tidal timescale estimates as
supplementary argumentation solely. We explore location of the
known planetary satellites on the ($\omega_0$, $e$) stability
diagram, where $\omega_0$ is an inertial parameter of a satellite
and $e$ is its orbital eccentricity. Using an empirical
relationship connecting the size of a satellite and its figure
asymmetry, we locate almost all known satellites on this diagram.
Then, by means of analysis of the residence of satellites in
various domains of stability/instability in this diagram, we draw
conclusions on the rotation states that are expected to be
predominant among the planetary satellites. Note that our
argumentation is independent from the postulates of the tidal
evolution theory: we judge on the possible rotation states solely
from the viewpoint of their expected stability or instability in
the current dynamical conditions, as inferred from observational
data.

\section{Synchronous resonance regimes}
\label{sec:regimes}

We consider the motion of a satellite with respect to its mass
centre under the following assumptions. The satellite is a
non-spherical rigid body moving in an elliptic orbit about a
planet. We assume the orbit to be fixed and the planet to be
a fixed gravitating point. These two assumptions are not
independent, because the flattening of the central planet leads to
precession of the orbits; see, e.g., \citep{RO97}. If one allows
for the precession of the orbit, e.g., if one considers the
non-sphericity of the planet as an ancillary perturbation, this
might lead exclusively (for generic values of the problem
parameters) to greater instabilities in the rotational motion. As
soon as our final result, stating it in advance, consists in the
statistical minority of the stable 1:1 solution, the accounting
for ancillary perturbations would generically strengthen this
final inference.

Besides, we assume that synchronous rotation is planar: the
rotational axis of a satellite coincides with the axis of the
maximum moment of inertia of the satellite and is orthogonal to
the orbit plane. This is just the rotational state the stability
of which we shall explore. The background for this assumption is
that the overwhelming majority of the satellites known to rotate
synchronously (all except the Moon) have the spin axis almost
orthogonal to the orbit plane \citep{SAA07}. Of course, this state
is expected from the tidal evolution theory \citep{GP66, P69,
P77}, but we take the predominance of planar rotation in
synchronous state solely as an observational fact. In what
concerns the collinearity of the rotational axis with the axis of
the maximum moment of inertia, the timescale of damping to this
state is very short \citep{P77}.

The shape of the satellite is described by a triaxial ellipsoid
with the principal semiaxes $a > b > c$ and the corresponding
principal central moments of inertia $A < B < C$. The dynamics of
the relative motion in the planar problem (i.e., when the
satellite rotates/librates in the orbital plane) are determined by
the two parameters: $\omega_0=\sqrt{3(B-A)/C}$, characterizing the
dynamical asymmetry of the satellite, and $e$, the eccentricity of
its orbit. Under the given assumptions, the planar
rotational--librational motion of a satellite in the
gravitational field of the planet is described by the Beletsky
equation~\citep{B59, B65}:
\begin{equation}
(1 + e \cos f){{\rm d}^2 \theta \over {{{\rm d} f}^2}} - 2e \sin f
\, {{\rm d} \theta \over {{\rm d} f}} + \omega_0^2 \sin{\theta}
\cos{\theta} = 2e \sin f,
\label{bel_eq}
\end{equation}
\noindent
where $f$ is the true anomaly, $\theta$ is the angle
between the axis of the minimum principal central moment of
inertia of the satellite and the ``planet -- satellite'' radius
vector.

Note that the traditional equation for the orientation of the
satellite as a function of time is
\begin{equation}
{{\rm d}^2 \vartheta \over {{{\rm d} t}^2}} + {{G M \omega_0^2}
\over {2r^3}} \sin{2(\vartheta - f)} = 0 ,
\label{dan_eq}
\end{equation}

\noindent \citep{B59, B65, C66, WPM84}, where $G$ is the
gravitational constant, $M$ is the mass of the planet, $r$ is the
module of the ``planet -- satellite'' radius vector, the angle
$\vartheta = \theta + f$ describes the orientation of the
satellite in an inertial coordinate system: it is the angle
between the axis of the minimum principal central moment of
inertia of the satellite and the planet--pericentre line.
Transformation to the Beletsky form (\ref{bel_eq}) is given, e.g.,
in \citep{B65, RO97}. The Beletsky form is convenient for further
numerical analysis at arbitrary eccentricities, including large
ones.

An analysis of Equation~(\ref{bel_eq}) by \citet{T64} showed that,
at certain values of the parameters the equation has two stable
$2\pi$-periodic solutions, i.e., there are two different modes of
rotation that are 1:1 synchronous with the orbital motion.
\citet{Z64} determined the boundaries of the stability domains of
these solutions in the ($\omega_0^2$, $e$) plane.
\citet{WPM84} noted the existence of these two different types of
synchronous resonance in application to results of their numerical
simulations of the rotation of Hyperion.

Let us recall the notions of these two kinds of synchronous 1:1
resonance, following~\citep{MS00}. For a satellite in an eccentric
orbit, at definite values of the $\omega_0$ inertial parameter,
synchronous resonance can have two centres in spin-orbit phase
space; in other words, there can be two different synchronous
resonances, stable in the planar rotation problem. Consider a
section, defined at the orbit pericentre, of the spin-orbit phase
space. At $\omega_0 = 0$, there exists a sole centre of
synchronous resonance with coordinates $\theta=0 \mbox{ mod }
\pi$, ${\rm d}\theta/{\rm d}t = 1$. If the eccentricity is
non-zero, upon increasing the value of $\omega_0$, the resonance
centre moves down the ${\rm d}\theta/{\rm d}t$ axis, and at a
definite value of $\omega_0$ another synchronous resonance
appears; at small eccentricities this value of $\omega_0$ is close to
1, see \citep[fig.~2]{MS00}, \citep[fig.~3]{MS08}.
Following~\citep{MS00}, we call the former resonance (emerging at
zero value of $\omega_0$) the {\it alpha} mode, and the latter
one~--- the {\it beta} mode of synchronous resonance. Upon
increasing the $\omega_0$ parameter, the alpha and beta modes
coexist over some limited interval of $\omega_0$ (the extent of
this interval depends on the orbital eccentricity), and in the
phase space section there are two distinct resonance centres
situated at one and the same value of the satellite's orientation
angle. For illustration see the phase space section in
Fig.~5c, given below. Such an effect takes place for
Amalthea~(J5)~\citep{MS98, MS00}; in~\citep{KS06} the conditions
for this effect, called there the ``Amalthea effect'', were
considered and discussed. On further increasing the $\omega_0$
parameter, at some value of $\omega_0$
the alpha resonance disappears, i.~e., it becomes unstable in the
planar problem, and only the beta resonance remains; for
illustration see \citep[figs.~1 and 2]{MS00},
\citep[fig.~3]{MS08}.

\section{The (\mbox{\boldmath{$\omega_0$}}, \mbox{\boldmath{\lowercase{$r$}}}) relationship}
\label{sec:relationship}

To make inferences on the possible rotational dynamics of a
satellite, one should know, in particular, its inertial
parameters, generally derived from the three-dimensional form of
the satellite. Such information is available now only for a very
limited number of satellites (less than 40). So, one has to find
ways of estimating these parameters from more available
characteristics, e.g, rough estimates of size. \citet{KS06} made
exponential and power-law fits to the dependences of the inertial
parameters $A/C$ and $B/C$ on the satellite radius $r$, defined as
the geometric mean $r = (abc)^{1/3}$ of the semiaxes of the
triaxial ellipsoid approximating the shape of the satellite. They
found that the exponential fits were better, having greater values
of the correlation coefficient. \citet{MS07} constructed the
dependence of the $\omega_0$ parameter on the satellite size
(radius) $r$. Following~\citep{KS06}, they fitted the statistical
($\omega_0$, $r$) relationship for 34 satellites by an
exponential function:
\begin{equation}
\omega_0(r) = A_0 \exp(-r/r_0),
\label{omega0_r}
\end{equation}
\noindent

\noindent and found $A_0 = 0.88 \pm 0.07$, $r_0 = 270 \pm 65$~km,
while the correlation coefficient $R^2 = 0.77$. Fig.~1 shows the
derived statistical dependence of the $\omega_0$ parameter on the
satellite size $r$. Approximation~(\ref{omega0_r}) is shown by the
solid curve.

Let us apply a simple power-law fit of the form $\omega_0(r) =
(r/r_0)^{-p}$ to the same data. This gives $p = 0.33 \pm 0.06$,
$r_0 = 7.22 \pm 2.59$~km, and a much smaller value of the
correlation coefficient: $R^2 = 0.58$; i.e., the power-law fit is
much worse. Also it is worse from the physical viewpoint: an
analysis of the expected form of the small satellites \citep{KS06}
predicts that in the limit $r \to 0$ the $\omega_0$ parameter
should be $\sim 1$, while the power-law fit gives infinity in this
limit; what is more, the decay of the fitting function at large
values of $r$ should be fast enough to describe the practically
zero values of $\omega_0$ for the big satellites, while the
power-law fit gives substantially non-zero $\omega_0$ values for
them.

So, exponential fit (\ref{omega0_r}) is much better and we use it
in what follows. Note that we introduce the continuous fit for
purely technical purposes, just to predict the $\omega_0$ value
straightforwardly, solely on the basis of an observational
estimate of the radius of a satellite. Does this continuous
function have any physical meaning? This depends on whether there
is a continuous transition in the shape asymmetry and size from
small to large satellites. There is an indication that there is no
continuous transition. As delineated in Fig.~1 by two dashed
rectangular boxes, the satellites can be roughly divided into two
groups: small and irregularly shaped (those with $r < 300$~km and
$\omega_0 > 0.2$; box~A) and big and round (those with $r >
500$~km and $\omega_0 < 0.2$; box~B). This division is highly
distinctive: there are no satellites in the range of radii $r$
from 300 to 500~km. What is more, the two groups are sharply
separated in $\omega_0$: there are no known small satellites with
$\omega_0 < 0.24$ and big satellites with $\omega_0 > 0.21$. Note
that, if we consider the primary inertial parameters $A/C$ and
$B/C$ instead of $\omega_0$, the satellites are not so well
separated: the small and big satellites overlap in the values of
them, see fig.~3 in \citep{KS06}.

It is remarkable that the division into two box populations
in Fig.~1 does not straightforwardly coincide with any known
physical division of the satellites, such as regular--irregular,
differentiated--undifferentiated,
impact-formed--hydrostatic-equilibrium divisions. Indeed, all
satellites in the diagram (in both boxes A and B), except Phoebe
in box~A and Triton in box~B, are regular; see satellite
classification and definition of regular and irregular satellites
in \citep{Sh06}. Box~A includes both differentiated and
undifferentiated and both impact-formed and
hydrostatic-equilibrium objects.

Note that, by pointing out the apparent division of the satellites
into two separate groups the ($\omega_0$, $r$) diagram, we do not
try to attach any physical or cosmogonical meaning to this
division. In fact, any physical mixture in box~A or B, or in the
sample as a whole, is not important for our dynamical study,
because only the moments of inertia and orbital eccentricity play
role in the equations.

Notwithstanding the possible separation of the data into two
groups, the continuous description (\ref{omega0_r}) is definitely
useful for statistical predictions, as testified by a rather high
value of the correlation coefficient and the over-all qualitative
agreement with the observed data, including the physically correct
behaviour in the limits $r \to 0$ and $r \to \infty$. We do
not explore the problem whether the exponential fit has any
physical meaning, as redundant for our further dynamical study.

\begin{figure}[h]
    \begin{center}
      \includegraphics{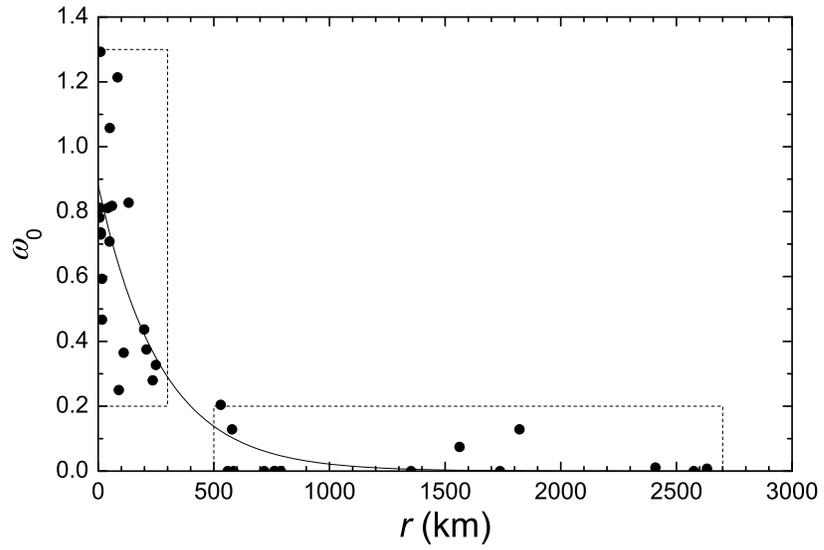}
    \end{center}
    \caption{Dependence of the $\omega_0$
parameter on the satellite radius $r$ (km). The solid curve
represents exponential approximation~(2) with $A_0 = 0.88$, $r_0 =
270$~km.}
    \label{fig1}
\end{figure}

{\bf\hspace{1cm}[Figure~\ref{fig1}]}

\begin{figure}[h]
    \begin{center}
      \includegraphics{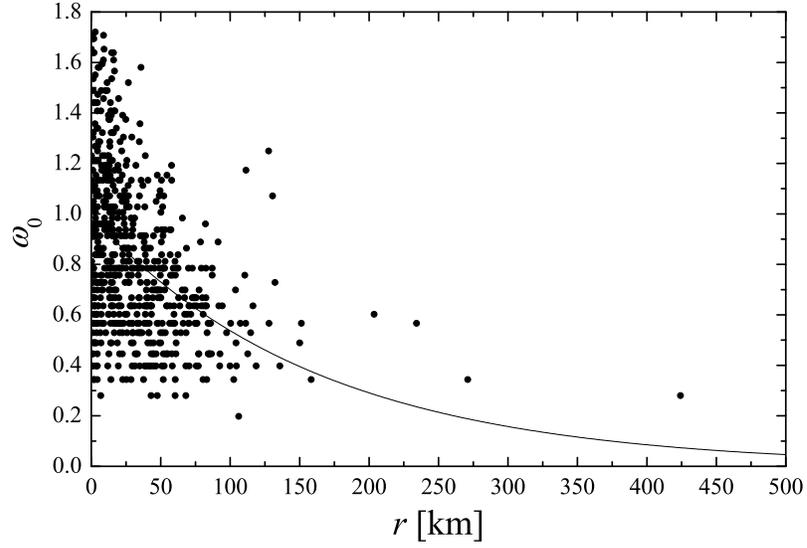}
      \includegraphics{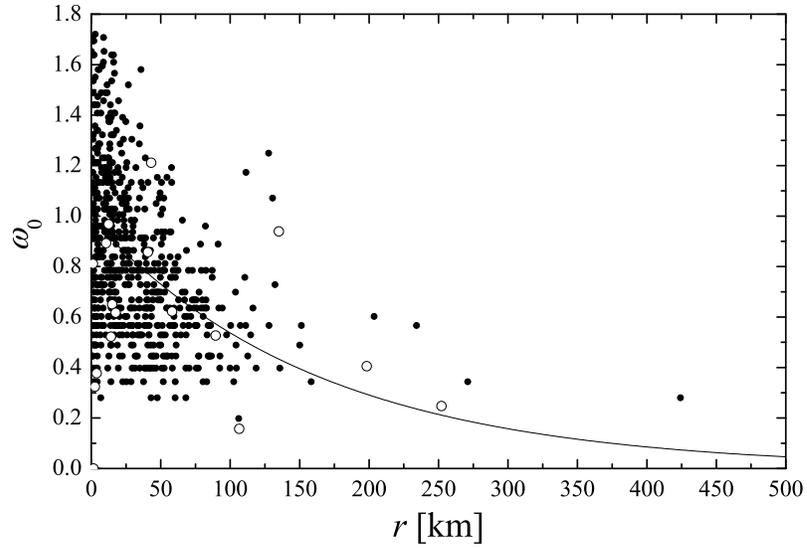}
    \end{center}
    \caption{(a) Dependence of the $\omega_0$
parameter on the asteroid radius $r$. The solid curve represents
exponential approximation~(2) with $A_0 = 0.99$, $r_0 = 163$~km.
(b) The same as in (a), but with the data on Saturnian satellites
\protect\citep{T10} superposed.}
    \label{fig2}
\end{figure}

{\bf\hspace{1cm}[Figure~\ref{fig2}]}

The fitting law~(\ref{omega0_r}) was derived in \citep{MS07} on
the basis of a very limited sample (including only 34 objects), so
the problem of its universality remains. We can try to acquire a
firmer confidence in it analyzing a similar relationship for
asteroids. Indeed, many planetary satellites can be a product of
orbital capture of asteroids~\citep{Sh06, JH07, J09}. Much greater
statistics (though of less quality) are available for asteroids
than for satellites, as we demonstrate below. We build the
statistical ($\omega_0$, $r$) relationship for asteroids in the
following way.

Let us rewrite the formula $\omega_0=\sqrt{3(B-A)/C}$ in the form
\begin{equation}
\omega_0 = \sqrt{3\frac{\gamma^2 - 1}{\gamma^2 + 1}} ,
\label{omega0_gamma}
\end{equation}

\noindent where $\gamma = a/b$, and $a > b > c$ are the semiaxes
of a homogeneous triaxial ellipsoid approximating the form of an
asteroid or a satellite. Representation~(\ref{omega0_gamma}),
valid for such an ellipsoid, follows from the formulas for the
moments of inertia expressed through the semiaxes; see, e.g.,
\citep{KS05}. From~(\ref{omega0_gamma}) it follows that {\it 1)}
$\omega_0$ depends only on $a$ and $b$, and does not depend on
$c$, {\it 2)} the upper limit for $\omega_0$ is equal to $\sqrt{3}
\approx 1.732$.

The amplitude $\Delta m$ of variation of the stellar magnitude of
an asteroid is given by the formula
\begin{equation}
\Delta m = - \frac{5}{4} \log \left( \cos^2 \varphi + \gamma^{-2}
\sin^2 \varphi \right) ,
\label{deltam}
\end{equation}

\noindent where $\varphi$ is the angle of the spin vector with
respect to the line of sight and it is assumed that the asteroids
are in a principal-axis rotation state; see
\citep{LL03,LJ07,MJD09}. Taking instead of $\varphi$ its mean
value $\bar \varphi$ (equal to 1) over hemisphere and inverting
formula~(\ref{deltam}), one obtains a relation for $\gamma$
through $\Delta m$:
\begin{equation}
\gamma = \sqrt{\frac{\sin^2 \bar \varphi}{10^{-\frac{4}{5}\Delta
m} - \cos^2 \bar \varphi}} \ ,
\label{gamtheta}
\end{equation}

\noindent Then it is substituted in formula~(\ref{omega0_gamma}).
So, $\omega_0$ can be approximately estimated if $\Delta m$ is
known.

Taking the tables from \citep{CALL} as a source of data on $\Delta
m$, in this way we estimate $\omega_0$ for 681 asteroids. The data
for trans-neptunian objects have been excluded from the sample,
since these data are strongly biased to large objects. (What is
more, TNOs might be physically irrelevant for comparing with most
of the satellites, see \cite{Sh06}.) Fig.~\ref{fig2}a shows the
derived statistical dependence of the $\omega_0$ parameter on the
asteroid radius $r$. Fitting the dependence by
formula~(\ref{omega0_r}) gives $A_0 = 0.99 \pm 0.02$, $r_0 = 163
\pm 15$~km; the correlation coefficient $R^2 = 0.18$. The fitting
curve is drawn in the Figure.

Comparing Figs.~\ref{fig1} and \ref{fig2}a, one can see that all
asteroids in Fig.~\ref{fig2}a except Ceres are situated inside the
boundaries of box~A of Fig.~\ref{fig1}, and none of the asteroids
correspond to population in box~B. Nevertheless the results of the
exponential fitting for the asteroids are very similar to those
for the whole sample of satellites. Though, as expected, the
correlation coefficient is much less for the asteroidal data, the
parameter values are similar, especially in the case of $A_0$.
Note that $A_0 = 1$ yields a physically justified limit $\omega_0
= 1$ for a small object size: this limit is consistent with the
expected axial ratios for a monolithic rock fragment, see
discussion in \citep{KS06}. It is remarkable that such satellites
(with $\omega_0 \approx 1$) in low-eccentricity orbits are subject
to the Amalthea effect (see Section~\ref{sec:regimes}); so the
fitting results predict that this effect should be usual for small
planetary satellites in close-to-circular orbits; of course, this
concerns only tidally-evolved objects, such as most of small
regular satellites and big ``particles'' of planetary rings.

For our further analysis it is important that the fitting
results for the asteroids provide a significant supplementary
justification of applicability of relationship~(\ref{omega0_r})
for estimating the expected value of $\omega_0$, when solely the
size of an object is available.

Let us return to the satellites. Up to now, when constructing
the ($\omega_0$, $r$) diagram, we have used an inhomogeneous
sample of objects. The available data for any separate homogeneous
sample of objects (e.g., satellites of a separate planet) are too
limited for statistical conclusions. However, one can make visual
inferences in the case of the Saturnian satellites, where the data
are most representative. In Fig.~2b, we superpose the latest data
about 17 Saturnian satellites, taken from \citep{T10}, on the
asteroidal data. We include only box~A satellite population, since
all (except one) asteroids fit in this box. One can see that the
points are even closer to the exponential curve (fitting the
asteroidal data) than many asteroids.

\section{The (\mbox{\boldmath{$\omega_0$}}, \mbox{\boldmath{\lowercase{$e$}}}) diagram}
\label{sec:diagram}

The ($\omega_0$, $e$) stability diagram is presented in
Fig.~3a. Theoretical boundaries of the zones of existence (i.e.,
stability in the planar problem) of synchronous resonances are
drawn in accordance with~\citep{M01}. Regions marked by ``Ia'' and
``Ib'' are the domains of sole existence of alpha resonance,
``II'' is the domain of sole existence of beta resonance, ``III''
is the domain of coexistence of alpha and beta resonances, ``IV''
is the domain of coexistence of alpha and period-doubling
bifurcation modes of alpha resonance, ``V'' is the domain of
non-existence of any 1:1 synchronous resonance, ``VI'' is the
domain of sole existence of period-doubling bifurcation modes of
alpha resonance. Domain V in Fig.~3a is not shaded, so we call it
henceforth the ``white domain''.

The structure of the diagram in the given ranges of $\omega_0$ and
$e$ in Fig.~3a is formed by four curves.

At the point $\omega_0 = 1$, $e = 0$, the branching curve is born
\citep{Z64,T64}. To the left-hand side from the branching curve,
only one uneven $2\pi$--periodic solution of the Beletsky equation
(alpha resonance) can exist; to the right-hand side, one or two
stable solutions (beta resonance or both alpha and beta
resonances, respectively) can exist \citep{Z64,T64}. (The
synchronous resonance corresponding to the stable uneven
$2\pi$--periodic solution existing only for $\omega_0 \ge 1$ is
beta resonance. The synchronous resonance corresponding to a
different stable uneven $2\pi$--periodic solution existing both at
$\omega_0 \ge 1$ and at $\omega_0 < 1$ is alpha resonance.)

At the point $\omega_0 = 1/2$, $e = 0$, the zone of parametric
resonance emerges. (It is ``v-shaped'' in the vicinity of the
point.) The left boundary of this zone corresponds to the loss of
stability of the uneven $2\pi$--periodic solution (alpha resonance
existing in domain~Ia) through the period-doubling bifurcation.
This line is the left dashed curve in Fig.~3a. The second (right)
dashed curve, drawn in accordance with \citep{M01}, corresponds to
the loss of stability of the bifurcated mode through the second
consecutive bifurcation. Thus the dashed borders of domains IV and
VI are formed by the lines of the first doubling bifurcation (on
the left) and the second doubling bifurcation (on the right).

The borders between zones have been found by means of a numerical
method of consecutive iterations. It is based on locating the
centre of synchronous resonance in the phase space section defined
at the pericentre of the orbit. As an initial approximation, one
takes the coordinates of the centre of resonance, known beforehand
(e.g., at $ \omega_0 = 0$ and $e \ge 0$ one has $ \theta = 0$ and
$d\theta/dt = 1$ for the centre of synchronous resonance). By
solving the Beletsky equation numerically, one finds the maximum
deviation (in the phase space section) of the trajectory from the
initial data. Taking the half of this deviation, one finds the
next approximation for the initial data. The iterative procedure
is stopped when the deviation is found to be zero, up to an
adopted level of accuracy. On varying the value of $\omega_0$ or
$e$, the moment of loss of stability (the moment of bifurcation)
is fixed, when the consecutive iterations for determining the
coordinates of the centre of resonance cease to converge, i.e.,
when the regular island corresponding to a given kind of
synchronous resonance ceases to exist. The obtained locations of
the branching curve and the borders of parametric resonance are in
agreement with graphs in \citep[fig.~4]{Z64} and
\citep[fig.~14]{B02}, where they were obtained by different
semi-analytical methods.

Graphical illustrations of the appearance and locations of various
kinds of resonances in phase space sections are given below in
Section~\ref{sec:kinds}.

\begin{figure}[h]
    \begin{center}
      \includegraphics{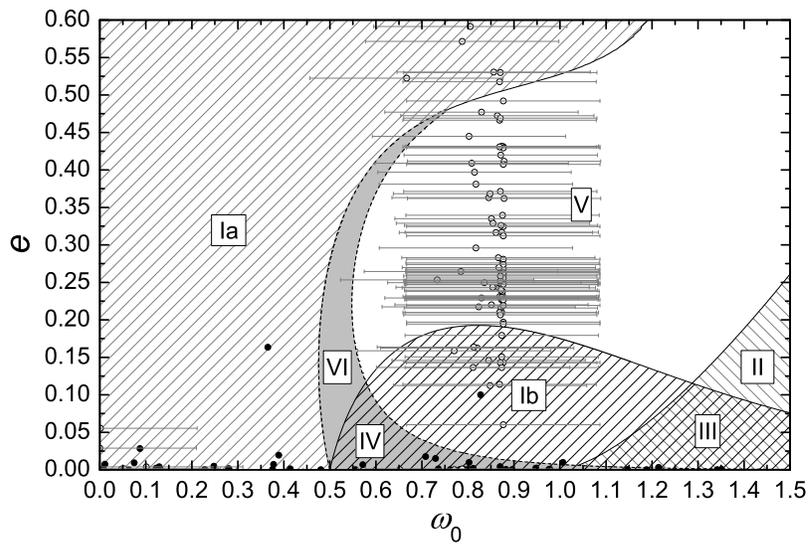}
      \includegraphics{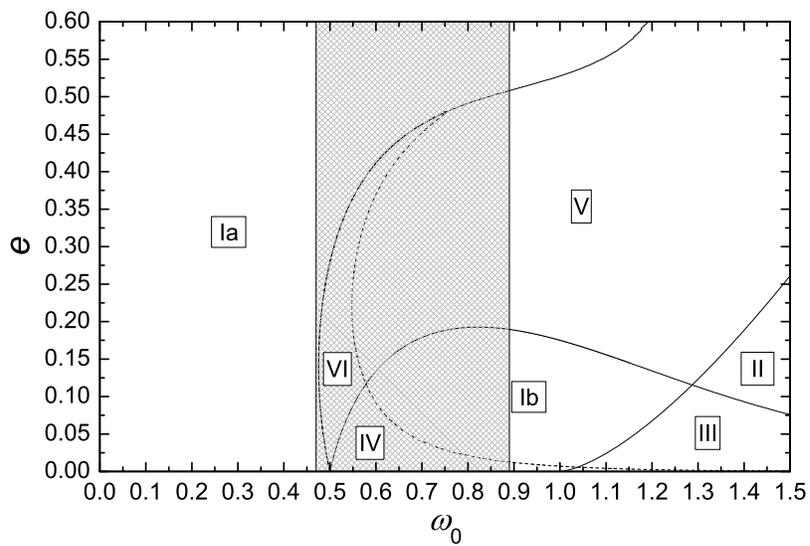}
    \end{center}
    \caption{(a) Location of the satellites
with known radii in the ($\omega_0$, $e$) diagram. (b) Predicted
location of the box~A population in the same diagram (the shaded
area).}
    \label{fig3}
\end{figure}

{\bf\hspace{1cm}[Figure~\ref{fig3}]}

To place satellites in the diagram, one should know the values of
$\omega_0$ and $e$. The values of $\omega_0$ are available now for
34 satellites only; see compilation and references
in~\citep{KS05}. For all other satellites (with unknown values of
$\omega_0$), following an approach proposed in \citep{MS07}, we
estimate $\omega_0$ by means of approximation~(\ref{omega0_r}) of
the observed dependence of $\omega_0$ on the satellite size $r$.
The data on sizes (and on orbital eccentricities) we take
from~\citep{K03, SJ03, SJK05, SJK06, P07, T07, NASA09b}.

In total, the data on sizes and orbital eccentricities are
available for 145 satellites. So, there are 145 ``observational
points'' in the ($\omega_0$, $e$) stability diagram in
Fig.~3a. (Note that, for other purposes, the locations of 10
satellites in the ($\omega_0$, $e$) diagram were
investigated in \citep{M01}, the position of 13 satellites in this
diagram was described in \citep{B02}, and the locations of 87
satellites in this diagram were studied in \citep{MS07}.) The
solid circles in Fig.~3a represent the satellites with known
$\omega_0$. The open circles represent the satellites with the
$\omega_0$ parameter determined by formula~(\ref{omega0_r}). The
horizontal bars indicate the three-sigma errors in estimating
$\omega_0$. They are all set to be equal to the limiting maximum
value $0.21$, following from the uncertainty in $A_0$. This gives
an approximate range of the possible values of $\omega_0$ at small
values of $r$.

From the constructed diagram we find that 73 objects are situated
in domain V (``white domain''). In domain Ib, 12 objects are
situated above Hyperion (a sole solid circle in domain Ib), while
two objects are below Hyperion. No synchronous states of rotation
exist in domain V. In the next Section we show that for the
majority of the satellites in domain Ib (namely, for the objects
above Hyperion) synchronous rotation is highly probable to be
attitude unstable. So, 73 objects in domain V and 12 objects in
domain Ib rotate either regularly and much faster than
synchronously (those tidally unevolved), or chaotically (those
tidally evolved). Summing up the objects, we see that a major part
(at least 85 objects) of all satellites with unknown rotation
states (132 objects), i.e., at least 64 per cent, cannot rotate
synchronously.

Note that the exponential (or any other) curve fitting of the
($\omega_0$, $r$) dependence is not crucial for achieving the
final results of our study. The curve fitting is used here solely
on technical reasons. Practically the same results can be achieved
without making curve approximations, but basing solely on the
division of the sample in the ($\omega_0$, $r$) plane into two
groups (box~A and box~B populations). This is demonstrated in
Fig.~3b, where the borders of the box~A population are superposed
on the stability diagram. These borders have been found without
making any curve fit, but solely by calculating the $\omega_0$
scatter for the objects in box~A. The shaded area in Fig.~3b
corresponds to the $\omega_0$ values in the range $0.68 \pm 0.21$.
This is the average value of $\omega_0$ taken with its $3 \sigma$
uncertainty for the satellites in box~A in Fig.~1. One can see
that this straightforward ``projecting'' of box~A into the
($\omega_0$, $e$) diagram predicts location for the satellites
with unknown inertial parameters similar to that given by the
exponential fitting procedure. However, the latter procedure is
definitely better, because it predicts the $\omega_0$ value in the
limit $r \to 0$, while the former uses the average $\omega_0$
value taken at $r$ from 0 to $\approx 250$~km.

\section{Attitude stability of synchronous rotation in domain I\lowercase{b}}
\label{sec:stability}

An analysis of the attitude stability of synchronous rotation
allows one to increase the expected number of satellites that
cannot rotate synchronously to an even greater value. To
demonstrate this, let us consider the attitude stability of the
satellites in domain Ib, i.e., the satellites in exact alpha
resonance.

The system of equations in variations with respect to the periodic
solution in this problem consists of six linear differential
equations of the first order with periodic coefficients. (The
original system of the Euler equations, describing the
three-dimensional rotational motion, is given, e.g., in
\citep[eqs.~6]{WPM84}. The Euler equations are linearized in the
variations; this gives the mentioned linear system.) Numerical
integration of the system allows one to obtain the matrix of
linear transformation of variations for a period; see
\citep{WPM84}. The periodic solutions in the given problem are
characterized by three pairs of multipliers. Following
\citep{MS00}, we build the distributions of the modules of
multipliers for a set of trajectories corresponding to a centre of
synchronous resonance on a grid of values of the $b/a$ and $c/b$
parameters. Analysis of the distributions allows one to separate
orbits stable with respect to tilting the axis of rotation from
those which are attitude unstable~\citep{MS00}.

\begin{figure}[h]
    \begin{center}
      \includegraphics[width=120mm]{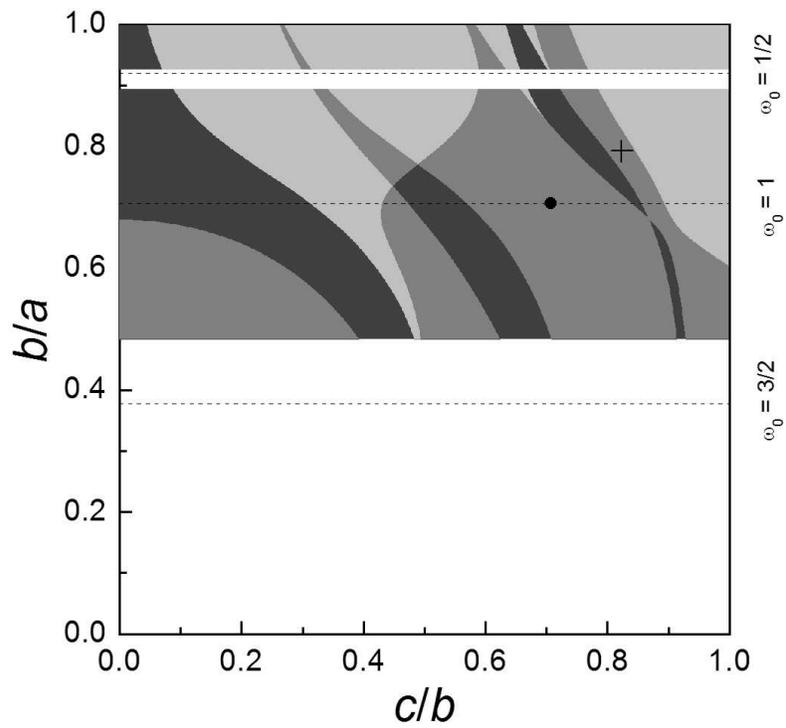}
    \end{center}
    \caption{Regions of stability and instability
with respect to tilting the axis of rotation; $e = 0.1$ (the
Hyperion case).}
    \label{fig4}
\end{figure}

{\bf\hspace{1cm}[Figure~\ref{fig4}]}

The computed regions of stability and instability are shown in
Fig.~4 for $e = 0.1$ characteristic for the Hyperion case. The
regions of stability are represented in light gray, the regions of
minimum (one degree of freedom) instability are in dark gray, and
the regions of maximum (two degrees of freedom) instability are in
black. The white areas in Fig.~4 correspond to case when
alpha resonance does not exist. (The intervals of values of
$\omega_0$, corresponding to the alpha resonance non-existence,
can be found from Fig.~3a, where they are given by the extents (in
$\omega_0$) of domains Ia and Ib at the fixed eccentricity
$e=0.1$.) Lines of constant values of the $\omega_0$ parameter
are drawn in Fig.~4 for reference. The location of Hyperion is
shown by a cross; the data on its $a$, $b$, and $c$ are taken
from~\citep{BNT95}. A bold dot represents the expected location of
a satellite with size tending to zero (see \cite{KS06}): $b/c =
b/a = 0.708$. Both the cross and the dot are apparently situated
in regions of instability, which occupy large portions of the area
of the diagram. With increasing $e > 0.1$, the area of the
instability regions only increases, and this means that for the
satellites situated in domain Ib above Hyperion there is almost no
chance to reside in an attitude-stable rotation state.

There are 15 objects present in domain Ib in Fig.~3a. Twelve
of them have orbital eccentricities greater than $0.1$. Therefore
one can add these 12 satellites to the total sample of the
satellites that are not expected to rotate synchronously.

\section{Basic kinds of phase space sections}
\label{sec:kinds}

To provide graphical illustrations to our conclusions, we
construct representative phase space sections of the planar
rotational motion in the basic domains in the ($\omega_0$, $e$)
stability diagram. The phase space sections are defined at the
pericentre of the orbit; i.e., the motion is mapped each orbital
period.

Basic qualitative kinds of the phase space sections, corresponding
to domains Ia, Ib, III, and IV, are presented in Fig.~5. For each
domain we take a representative satellite. Namely, the sections
are constructed for Phoebe~(S9) ($e = 0.176$, $\omega_0 = 0.365$;
thus belonging to domain Ia), Hyperion~(S7) ($e = 0.100$,
$\omega_0 = 0.827$; domain Ib), Amalthea~(J5) ($e = 0.003$,
$\omega_0 = 1.214$; domain III), and Pandora~(S17) ($e = 0.004$,
$\omega_0 = 0.870$; domain IV).

\begin{figure}[h]
    \begin{center}
      \includegraphics[width=80mm]{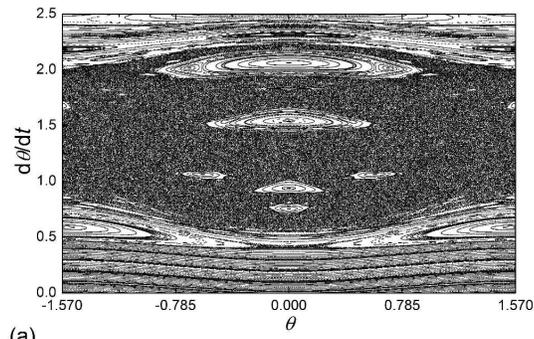}\\
      \includegraphics[width=80mm]{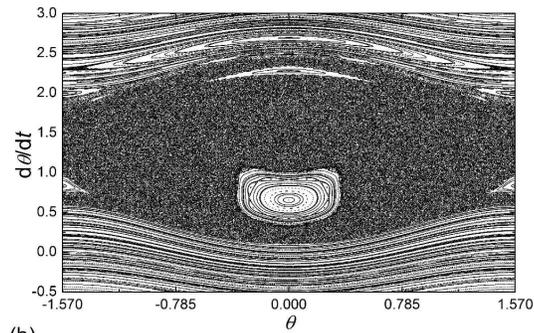}\\
      \includegraphics[width=80mm]{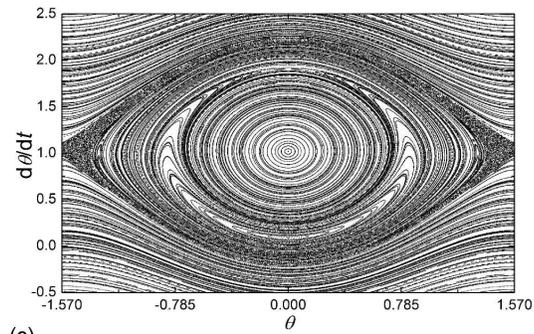}\\
      \includegraphics[width=80mm]{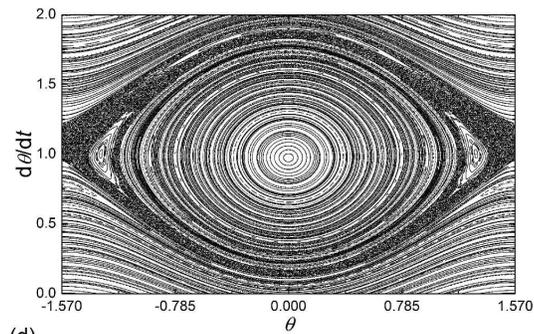}\\
    \end{center}
    \caption{Basic qualitative kinds of the phase
space sections for the planar rotational motion. (a) Phoebe
(domain Ia), (b) Hyperion (domain Ib), (c) Amalthea (domain III),
(d) Pandora (domain IV).}
    \label{fig5}
\end{figure}

{\bf\hspace{1cm}[Figure~\ref{fig5}]}

In the cases of domains Ia and Ib, the phase space sections
contain a broad chaotic layer with alpha resonance inside
(Figs.~5a, b). In the case of domain III (Fig.~5c), there exist
alpha resonance (the lower one in the section) and beta resonance
(the upper one). In the case of domain IV, there exist alpha
resonance and its period-doubling bifurcation mode. The latter
mode appears as the two islands inside the chaotic layer, to the
left and to the right of alpha resonance (Fig.~5d).

\begin{figure}[h]
    \begin{center}
      \includegraphics[width=80mm]{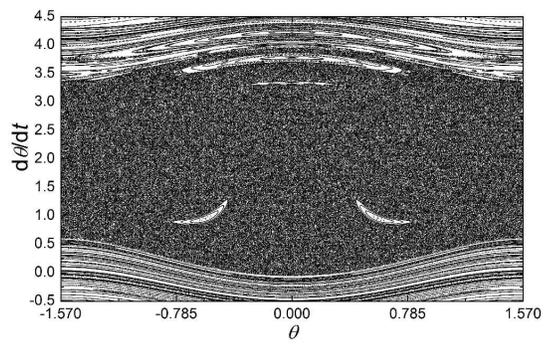}
    \end{center}
    \caption{The phase space section for the
planar rotational motion of an object in the ``white domain''.}
    \label{fig6}
\end{figure}

{\bf\hspace{1cm}[Figure~\ref{fig6}]}

{\bf\hspace{1cm}[Table~1]}

A representative phase space section in the case of domain V
(``white domain'') is given in Fig.~6. Here we use model values of
the parameters, namely, $e = 0.25$ and $\omega_0 = 0.9$. They
roughly correspond to the centre of the ``white domain''. No
synchronous state (neither alpha nor beta) exists in the phase
space section. There is a prominent chaotic sea instead.

A practical guide for interpretation of details of the phase space
sections in Figs.~5 and~6 is provided by Table~1, where prominent
regular islands in the sections are identified with resonances.

\section{Despinning times}
\label{sec:times}

If the orbit is fixed, the planar rotation (i.e., the
rotation with the spin axis orthogonal to the orbital plane) in
synchronous 1:1 resonance with the orbital motion is the most
likely final mode of the long-term tidal evolution of the
rotational motion of a planetary satellite \citep{GP66, P69, P77}.
In this final mode, the rotational axis of a satellite coincides
with the axis of the maximum moment of inertia of the satellite
and is orthogonal to the orbital plane. If the orbit exhibits
nodal precession, the typical final evolutionary state of the spin
axis of a satellite is a low-obliquity Cassini state \citep{C66,
P69, P77, C09}.

Calculation of the time of despinning to synchronous state,
due to tidal evolution, shows whether a satellite's spin could
evolve to synchronous state since the formation of the satellite.
For a satellite to be ultimately captured in synchronous 1:1 spin
state, or any other spin-orbit resonance, the current dynamical
and physical properties of the satellite should allow for a
sufficiently short, at least less than the age of the Solar
system, time interval of tidal despinning to the resonant state.

Let us consider theoretical estimates of the despinning time for
the satellites in the ``white domain'' and domain Ib. We estimate
the tidal despinning time of a satellite by means of the following
formula~\citep{D95}:
\begin{equation}
   T_\mathrm{despin} = \frac{\omega_\mathrm{I} - \omega}{|{\dot \omega}|},
   \label{eq:Tdespin}
\end{equation}

\noindent where $\omega_\mathrm{I}$ and $\omega$ are the initial
and the final spin rates of a satellite, respectively, and
\begin{equation}
   |{\dot \omega}| = \frac{45\rho r^2 n^4}{38\mu Q}
   \label{eq:K}
\end{equation}

\noindent is the absolute value of the rate of the rotation
slowdown (see equations~(1) and (11) in~\citep{D95}). Here $r$ is
the satellite radius, $n = 2\pi/T_\mathrm{orb}$ is its orbital
frequency (the mean motion), $\rho$ and $\mu$ are the density and
rigidity of the satellite, respectively; $Q$ is the satellite's
tidal dissipation function. Eq.~(\ref{eq:K}) corresponds to the
commonly considered case of the low orbital
eccentricities~\citep{D95}. In case of the high eccentricities we
use formula~(4) from~\citep{D95}:
\begin{equation}
   |{\dot \omega}| = (1-e^2)^{-9/2}\left(1 + 3e^2 +\frac{3}{8}e^4\right)
   \frac{45\rho r^2 n^4}{38\mu Q}.
   \label{eq:K_e}
\end{equation}

\noindent We estimate $T_\mathrm{despin}$ for two sets of the
values of parameters, namely, the set adopted and justified
in~\citep{D95} ($\omega_\mathrm{I} = 2\pi/(10\;\mathrm{hours})$,
$\rho = 1\;\mathrm{g}\,\mathrm{cm}^{-3}$, $\mu = 3.5 \times
10^{10}\;\mathrm{dyn}\,\mathrm{cm}^{-2}$) and the set adopted and
justified in~\citep{P77} ($\omega_\mathrm{I} =
2\pi/(2.3\;\mathrm{hours})$, $\rho =
2\;\mathrm{g}\,\mathrm{cm}^{-3}$, $\mu = 5 \times
10^{11}\;\mathrm{dyn}\,\mathrm{cm}^{-2}$). The final $\omega$ is
fixed to be equal to $2\pi/T_\mathrm{orb}$, i.~e., the value at
synchronous resonance, but note that for any estimates by the
order of magnitude the choice of the exact final $\omega$ does not
matter much.

The calculation of despinning times for the objects in domains V
and Ib shows that the minimum values of these times belong to
Elara (J7) with $6 \times 10^{12} Q$ years, Carme (J11) with $4
\times 10^{13} Q$ years, and Themisto (J18) with $7 \times 10^{12}
Q$ years. These minimum values are exceeded by typical ones in the
set by 2--3 orders of magnitude. Taking $Q \sim 100$~\citep{P77},
one finds that the despinning times of the satellites in domains V
and Ib are by far large in comparison with the Solar system age.
The spins of these satellites could not have evolved up to
entering the chaotic zone near low order spin-orbit resonances.
This is in agreement with the general conclusion by \citet{P77}
that most of irregular satellites still remain in spin states
close to initial ones.

The irregular satellites can be a product of orbital
capture~\citep{Sh06, JH07, J09} or disruption of larger
bodies~\citep{C94, Sh06, J09}. If the objects in domains V and Ib
are a product of recent capture or disruption, the allowed times
for evolution are even smaller. If most of them originated from
capture from the asteroidal population, the spin distribution
among the satellites should have remained practically unchanged.
Could a captured asteroid have a large enough initial rotation
period that allowed immediate entering the chaotic zone?

According to~\citep{PH00, HP06}, among the asteroids with known
rotation periods there exists a statistically distinct (about 2
per cent of the total population) group of ``slow rotators'' with
measured rotation periods up to $\approx 1000$ hours (50 days); an
observed lower boundary for the rotation periods of the asteroids
in this group is $\approx 30$~hours for the objects with diameter
less than 10~km. An even greater excess of slow rotators was found
in new surveys by \cite{PHV08} and \cite{MJD09}. \cite{PHV08}
explain this excess as due to the Yarkovsky effect.

In principle, a satellite with a small enough orbital period, like
Themisto (with $T_{\mathrm{orb}} \simeq 130$~d), if it were such
an outlier captured in the orbit, could have entered the chaotic
zone in phase space. One should take into account that the extent
of the chaotic zone, depending on the inertial and orbital
parameters, might be rather large in rotation frequency, the upper
border being an order of magnitude greater than the synchronous
frequency value, see figs.~7--9 in~\citep{D95} and fig.~3
in~\citep{KS05}. However, the chances are apparently low, and one
can hardly expect that more than one or two satellites in the
``white domain'' rotate chaotically,~--- of course, if one takes
for granted that the tidal processes are well understood, and so
the real tidal evolution could not be faster. An example provided
by Iapetus (S8) shows that there might be situations when the
tidal evolution is much faster than in the standard theories:
there is an inconsistency of at least two orders of magnitude
between the calculated (large) despinning time for this satellite
and the observational fact that it is tidally despun \citep{CR07,
A09}. Perhaps a reconsideration of the current tidal despinning
theories is needed \citep{EW09,G09}. The traditional tidal
approach is especially unreliable in the case of highly eccentric
and inclined orbits \citep{G09}, i.e., for the irregular
satellites. What is more, individual scenarios of orbital
evolution for irregular satellites can exist, in which the
despinning times are decreased radically, as demonstrated in
\citep{D95} for the case of Nereid (N2), the satellite with the
maximum known orbital eccentricity. All these facts, combined with
the opportunity of capture of slow rotators, demonstrate that the
tidal theory prediction that the current rotation of all irregular
satellites is fast might be a subject for revision.

\section{Conclusions}
\label{sec:concl}

On the basis of tidal despinning timescale arguments, Peale showed
in 1977 that the majority of irregular satellites are expected to
reside close to their initial (fast) rotation states. Here we have
investigated the problem of typical rotation states among
satellites from a purely dynamical stability viewpoint. We have
shown that, though the majority of the planetary satellites with
known rotation states rotate synchronously (facing one side
towards the planet, like the Moon), a significant part (at least
64 per cent) of all satellites with unknown rotation states cannot
rotate synchronously. The reason is that no stable synchronous 1:1
spin-orbit state exists for these bodies, as our analysis of the
satellites location in the ($\omega_0$, $e$) stability diagram
demonstrates. They rotate either regularly and much faster than
synchronously (those tidally unevolved) or chaotically (tidally
evolved objects or captured slow rotators).

With the advent of new observational tools, more and more
satellites are being discovered. Since they are all small, they
are all irregularly shaped, according to formula~(\ref{omega0_r}).
Besides, the newly discovered objects typically move in strongly
eccentric orbits~\citep{NASA09b, SJ03}. Therefore these new small
satellites are all expected to be located mostly in the ``white
domain'' of the ($\omega_0$, $e$) stability diagram,
i.e., the synchronous 1:1 rotation state is impossible for them.

\pagebreak

\renewcommand{\baselinestretch}{1}

\noindent{\large\bf Table~1}

\bigskip\noindent{\bf The centres of resonances in the phase space sections}

\begin{table}[h]
{\small
\begin{tabular}{lccccc}
\hline\noalign{\smallskip}
Resonance & Phoebe (S9)     & Hyperion (S7)       & Amalthea (J5) & Pandora (S17)        & Model (Fig.~6)  \\
\hline\noalign{\smallskip}
alpha & $(0, 0.940)$        & $(0, 0.659)$        & $(0, 0.090)$  & $(0, 0.976)$         & -- \\
beta  &         --          & --                  & $(0, 1.018)$  & --                   & -- \\
period-doubling \\
alpha &         --          & --                  & --            & $(\mp 1.302, 1.003)$ & -- \\
1:2   & $(\mp\pi/2, 0.589)$ & $(\mp\pi/2, 0.836)$ & --             & $(\mp\pi/2, 0.784)$  & -- \\
5:4   & --                   & --                   & --             & --                    &  $(\mp 0.587, 0.957)$\\
3:2   & $(0, 1.544)$        & --                   & --             & $(0, 1.901)$         & -- \\
9:4   & $(\mp\pi/2, 2.182)$,& $(\mp\pi/2, 2.000)$,& --             & --                    & -- \\
9:4   & $(0, 2.319)$        & $(0, 2.499)$        & --             & --                    & -- \\
2:1   & $(0, 2.063)$        & $(0, 2.280)$        & --             & --                    & -- \\
5:2   &          --          & $(0, 2.694)$        & --             & --                    & -- \\
7:4   & $(\mp\pi/2, 1.677)$ & --                   & --             & --                    & -- \\
3:1   & --                   & --                   & --             & --                    & $(0, 3.342)$\\
7:2   & --                   & --                   & --             & --                    & $(0, 3.782)$\\
4:1   & --                   & --                   & --             & --                    & $(0, 4.225)$\\
\noalign{\smallskip}\hline
\end{tabular}
\\

{\small The numbers in parentheses give the coordinates of the
centres of regular islands in Figs.~5 and~6. The ``--'' symbol
signifies that the corresponding mode is absent or is visually
unresolved in the section or is situated out of the borders of the
graph.} }
\end{table}

\pagebreak

\pagebreak

\pagebreak

\pagebreak

\pagebreak

\pagebreak

\end{document}